\documentclass[twocolumn,aps,superscriptaddress,prl]{revtex4}
\usepackage{amsmath}
\usepackage{amsfonts}
\usepackage{amssymb}
\usepackage{dcolumn}
\usepackage{bm}
\usepackage[sort&compress]{natbib}
\usepackage{ifpdf}
\ifpdf
   \usepackage[pdftex]{graphicx}
   \usepackage{graphics}
\else
   \usepackage{graphicx}
   \usepackage{graphics}
\fi
\title{Sylla_Manuscript}
\bibpunct{[}{]}{,}{n}{,}{,}

\begin{document}

\title{\Large{Anticorrelation between Ion Acceleration and Nonlinear Coherent Structures from Laser-Underdense Plasma Interaction}}

\author{F. Sylla}
\affiliation{%
Laboratoire d'Optique Appliqu\'ee, ENSTA, CNRS, Ecole Polytechnique, UMR 7639, 91761 Palaiseau, France\\
}%
\author{A. Flacco}
\affiliation{%
Laboratoire d'Optique Appliqu\'ee, ENSTA, CNRS, Ecole Polytechnique, UMR 7639, 91761 Palaiseau, France\\
}%
\author{S. Kahaly}
\affiliation{%
Laboratoire d'Optique Appliqu\'ee, ENSTA, CNRS, Ecole Polytechnique, UMR 7639, 91761 Palaiseau, France\\
}%
\author{M. Veltcheva}
\affiliation{%
Laboratoire d'Optique Appliqu\'ee, ENSTA, CNRS, Ecole Polytechnique, UMR 7639, 91761 Palaiseau, France\\
}%
\affiliation{%
Dipartimento di Fisica "G. Occhialini", Universit\`a degli Studi di Milano-Biccoca, piazza della Scienza 3, 20126 Milan, Italy\\
}%
\author{A. Lifschitz}
\affiliation{%
Laboratoire d'Optique Appliqu\'ee, ENSTA, CNRS, Ecole Polytechnique, UMR 7639, 91761 Palaiseau, France\\
}%
\author{G. Sanchez-Arriaga}
\affiliation{%
CEA, DAM, DIF, 91297 Arpajon, France\\
}%
\author{E. Lefebvre}
\affiliation{%
CEA, DAM, DIF, 91297 Arpajon, France\\
}%
\author{V. Malka}
\affiliation{%
Laboratoire d'Optique Appliqu\'ee, ENSTA, CNRS, Ecole Polytechnique, UMR 7639, 91761 Palaiseau, France\\
}
\date{\today}
\begin{abstract}
In laser-plasma experiments, we observed that ion acceleration from the Coulomb
explosion of the plasma channel bored by the laser, is prevented when multiple plasma instabilities
such as filamentation and hosing, and nonlinear
coherent structures (vortices/post-solitons) appear in the wake of an ultrashort laser pulse. The tailoring of the longitudinal plasma density ramp allows us to control the onset of these
insabilities. We deduced that the laser pulse is depleted into these structures in our conditions, when a plasma at about
$10\%$ of the critical density exhibits a gradient on the order of $250~\mu$m (gaussian fit), thus hindering the
acceleration. A promising experimental setup with a long pulse is demonstrated enabling the
excitation of an isolated coherent structure for polarimetric measurements and, in further perspectives, parametric studies of ion plasma
acceleration efficiency.  
\end{abstract}
\maketitle
\begin{figure}
\includegraphics[scale=1]{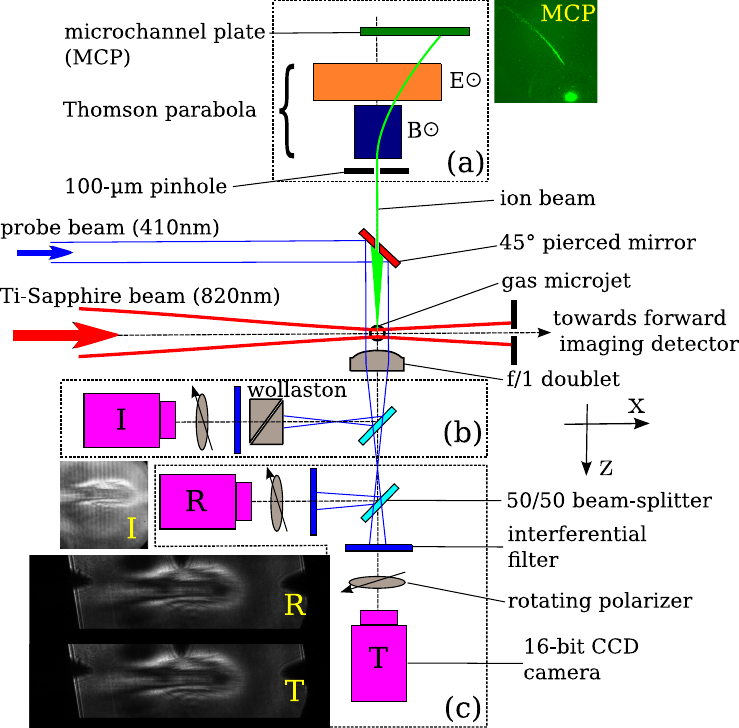}
\caption{Experimental setup for the interaction of a microjet and a short Ti-Sapphire pulse,
illuminated by a linearly-polarized-frequency-doubled probe pulse. (a) Imaging Ion spectrometer
along the normal to the
main beam with Thomson parabola ($B=0.3$~T) and chevron microchannel plate (MCP). A second ion
spectrometer is also set on the laser axis. (b) Nomarski interferometer. (c) Polarimeter using Faraday rotation effect. For Polarigrams R and T, the polarizers are rotated by respectively $+10$ and $-10^{\circ}$ fom the extinction}.
\label{setup}
\end{figure}

Laser-underdense plasma acceleration rests on the interaction of an intense relativistic pulse
($I\lambda^2>10^{18}$ Wcm$^{-2}\mu$m$^2$) with a plasma in a large range of electronic density
($n_e\in[10^{18},10^{20}]$ cm$^{-3}$) \cite{malk08,krus99,wei04,will06}. In particular, the Coulomb
explosion of a positively charged channel, created by the laser as it propagates in the plasma, is an
efficient way to radially accelerate ions \cite{sark99,krus99}. However, this acceleration can be accompanied by the copious generation of non-linear plasma structures that
significantly deplete the pulse energy, theoretically up to $30-40\%$ of the laser
energy \cite{bula99,sent99,naum01,esir02,esir08}. These structures aroused the interest of numerous
theoretical and experimental studies \cite{bula96,bula99,esir02,sanc11,sanc11a,borg02,roma10}. An underdense
plasma with $n_e\sim$ tens of percent of $n_c$, $n_c$ being the critical density, is particularly suitable for the excitation of coherent structures like electromagnetic solitons and
electron vortices, as predicted in Refs.~\cite{bula99,sent99,bula96,esir02} and observed with proton
radiography diagnostics in Refs.~\cite{borg02,sarr10,roma10}. In Ref.~\onlinecite{esir02}, a 3D
electromagnetic cavity structure in the wake of the pulse, termed
transverse magnetic-soliton, is identified and slowly expands into a
post-soliton, as ions within the cavity explode to energies of tens of keV. This example
illustrated a novel ion acceleration scheme from underdense plasma, different from the channel Coulomb explosion \cite{krus99},
Target Normal Sheath Acceleration-like process \cite{will06}, or longitudinal magnetic vortex acceleration \cite{kuzn01}.
\paragraph{}
In this letter, we present for the first time an ``inverse correlation'' (mutual exclusion) between the ion acceleration by
radial Coulomb explosion and the generation of nonlinear coherent structures in the wake of the
laser pulse. Our results tend to prove that the efficiency of the laser ion acceleration directly from a gas
jet is highly dependent upon the density gradient profile. In our case, though the density ramp
should contribute to the increase of the laser intensity by self-focusing of the pulse
\cite{sun87,mori88}, it induces filamentation \cite{mori88}, hosing instabilities \cite{spra94} and
ultimately generation of nonlinear coherent structures, the overall turning out to be detrimental for the acceleration.
\paragraph{}
The experiment was conducted at the Laboratoire d'Optique Appliqu\'ee with the Ti:Sapphire laser ``Salle
Jaune'', which delivers an ultrashort (duration $\tau=$ 30
fs), linearly-polarized (along $z$, see FIG.~\ref{setup}) pulse with $E_{L}\sim$810 mJ on target
\cite{flac10a}. The pulse was focused with
an $f/10$ off-axis parabolic mirror to a 25-$\mu$m ($1/e^{2}$) focal spot onto a 700 $\mu$m (FWHM)
supersonic He jet at $\sim$300 $\mu$m above the nozzle exit. The vacuum laser intensity is
$I_{0}\sim1\times10^{19}$ Wcm$^{-2}$. With $n_e=1-10\times10^{19}$ cm$^{-3}$, it gives a ratio
$P_L/P_c=5.3-53$ (with $P_L$ the laser power and $P_c=16.2$ $n_c/n_e$ GW the critical power for
self-focusing). Before focusing, the beam was split and $10\%$ of the energy sent onto a BBO crystal
for second harmonic generation. The intensity and spatial quality of this collimated
frequency-doubled beam were improved by passing through a $50$-$\mu$m-pinhole-filtered beam reducer.
It was finally incident on the jet normally to the main laser axis for illuminating and probing the
interaction transversally (see FIG.~\ref{setup}). Measurements of ion emission and plasma probing
(interferometry and polarimetry), both at $90^{\circ}$ from the main laser propagation axis, were simultaneously done by means of a
$45^{\circ}$-pierced mirror reflecting the probe beam onto the gas jet without cutting the ion
beam (see FIG.~\ref{setup}).
\paragraph*{}
The ion detector is composed of a 100-$\mu$m spatial filter (solid angle acceptance is $1.2\times10^{-8}$ sr) with a Thomson parabola spectrometer coupled to 40-mm imaging microchannel
plate chevron assembly (see FIG.~\ref{setup}(a)). A linearized 16-bit CCD camera
($1024\times1024$ pixels) enables single-shot recordings of scintillating traces and thus shot-to-shot statistical analyses of
ion spectra.
\paragraph*{}
An $f/1$ achromatic doublet coupled with a 50/50 beam splitter enables to record at the same time
with spatial resolution $\sim1~\mu$m, interferograms encoding the plasma density variations using a
Nomarski interferometer \cite{bena79} (see FIG.~\ref{setup}(b)), and polarigrams revealing the component of the
magnetic field collinear to the probe beam direction (see
FIG.~\ref{setup}(c)). The interferometer contains a Wollaston biprism of $10\times10$~mm$^{2}$ with
a $410$-nm interferential filter to improve the temporal coherence of the interferences. The
polarimeter is composed of two identical assemblies, each comprising in particular a Glan Laser
polarizer (extinction ratio better than $10^{-4}$) and a 16-bit CCD camera. For each assembly, a
scan of the rotating polarizer was carried out for probe beam transmission calibration (every five degrees
over the entire rotation). For the measurements, the
polarizers were detuned by a common angle from each minimum of transmission (typical detunings $\theta_R=+10$ and $\theta_T=-10^{\circ}$
for respectively the transmitted and reflected images from the beam splitter, see
FIG.~\ref{setup}(c)).
\paragraph*{}
The magnetic field detection rests on the measurment of the rotation, due to Faraday effect, of the
incident linear polarization
of the probe beam electric field when it crosses the plasma \cite{stam75,kalu10}. The
total rotation angle along a path $l_z$ is given by
$\phi_{rot}(x,y)=\frac{e}{2m_ecn_c}\int_{l_z}\!n_e(x,y,z)B_z dz$. Here $B_z$ is the magnetic field component collinear to the probe beam direction, $n_c$ the plasma critical density for the
probe beam, $e$ and $m_e$ the mass and charge of the electron and $c$ the light velocity in vacuum.
Taking the intensity ratio from the two shadowgrams R and T obtained with opposite detuning angles (see
FIG.~\ref{setup}(c)) prevents the extracted $\phi_{rot}$ map from being corrupted by probe beam
shot-to-shot fluctuations in intensity, and image shift from plasma refraction or vibrations. However,
this requires a pixel-to-pixel correspondance of the pictures. We achieved a sub-pixel-accurate
matching of the two pictures by tagging the field of view at three reference-areas prior to
the beam-splitting, and by running a numerical pattern matching routine.
\paragraph*{}
\begin{figure}
\includegraphics[scale=0.8]{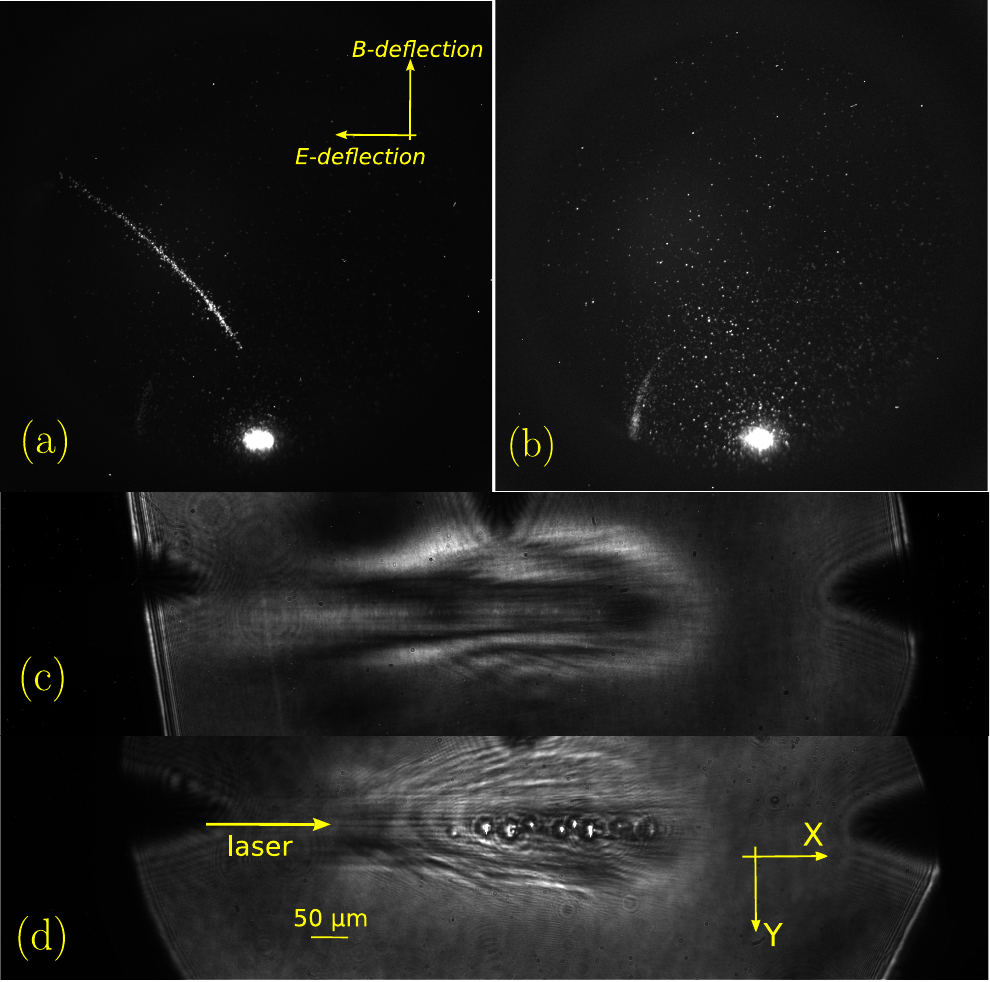}
\caption{(a) and (b): MCP recordings normally to the laser axis for $n_e\sim10^{20}$ cm$^{-3}$.
(a) short plasma gradient $\sim 150~\mu$m.
He$^+$ trace with energy cutoff $\sim250$ keV. (b) long plasma gradient $\sim 250~\mu$m. (c) and
(d): Shadowgram at 1 ps after the pulse enters the jet (resolution $\sim 1~\mu$m, magnification
8.2). (c) short plasma gradient. (d) long plasma gradient. The shadowgram shows a clear train of
bubble-like structures in the laser wake. The laser polarization is normal to
the picture (S-polarized)}
\label{trace}
\end{figure}
Figure~\ref{trace} shows the MCP recordings at 90$^{\circ}$ from the laser axis for two plasma gradient conditions. In (a), the laser is
shot on the gas jet at an early stage of its expansion in the vacuum chamber so that the plasma
gradient is steep $\sim 150~\mu$m (with a gaussian fit). In (b), the interaction occurs at a
late time of the expansion (80 ms after the closing of the valve), and the gradient is smooth $\sim
250~\mu$m. For both conditions, the peak density in the jet has the same value $n_e\sim10^{20}$
cm$^{-3}$. We observe \textit{systematically} radial ion acceleration from Coulomb explosion in case (a) and no trace in (b). In both cases, no ion was detected in the forward direction (see FIG.~\ref{setup} for the geometry). When
acceleration occurs, only He$^{+}$ are detected from 12 keV to cutoff energies of 250 keV. We attribute the
absence of He$^{2+}$, though the laser peak intensity is 1000 times higher than the helium
potential of full ionization, to efficient recombination of this species in the neutral corona
starting at $100 \mu$m from the laser axis. Indeed, the electron capture is the most probable
channel of interaction (cross section $\sigma\sim 10^{16}$ cm$^{-2}$) for He$^{2+}$ projectile of
250 keV streaming in He gas \cite{shah85}. In panel (d), one can see the typical shadowgram
(revealing sharp electron density fronts) obtained in case (b). A train of zig-zag ordered and well
separated bubble structures with an average size of about $30~\mu$m ($\gg\lambda_p=c/\omega_{pe}$, the plasma wavelength) can be clearly seen in the wake of the
laser pulse. The bubbles have a bright spot in their centers that is not observable when the
probe beam is turned off. This means that the emission is not of self-emission nature and results
from refractive effects of the probe light by these structures behaving like microlenses. An interesting
point is the quasi-circular shape of these structures, meaning that, during the transverse
illumination of 30 fs, their centers are quasi-static in the picture plane. In the short-plasma
scale length case, for which transverse ion acceleration is detected (Fig.~\ref{trace}(a)), no such
isolated structure is ever observed in the gas jet (Fig.~\ref{trace}(c)).
\paragraph*{}
To understand this inverse correlation, we investigated in time the interaction in the long gradient
case. Figure~\ref{filamentation} presents a sequence of five shadowgrams taken at different delays
after the pulse enters the jet ($t_0$). Very early in the interaction ($t_0+540$ fs), sizeable
filaments develop next to the main laser channel \cite{mori88} (see arrows in panel (b) from
Fig.~\ref{filamentation}). As the pulse propagates along, the filamentation becomes stronger and multiple circular structures appear about the laser channel, near the center of the jet.
In this time-resolved scan, we observe that the structures appear randomly in the laser path and no correspondance could be drawn between the
pulse and bubble positions. Moreover, even if this shadowgram diagnostic is not intrinsically
time-resolved, contrary to the
proton radiography technique \cite{roma05}, the stability of the interaction was sufficient to assess
statistically the average size of the structure until the moment when the shadowgrams become
blurred and unexploitable, at about 7 ps after the onset of the interaction. Over that timespan, we
observed that the structures have fairly the same size, irrespective of their positions to the laser
front and measured an average increase of less than $10~\mu$m in size.
\begin{figure}[h]
\includegraphics[scale=0.4]{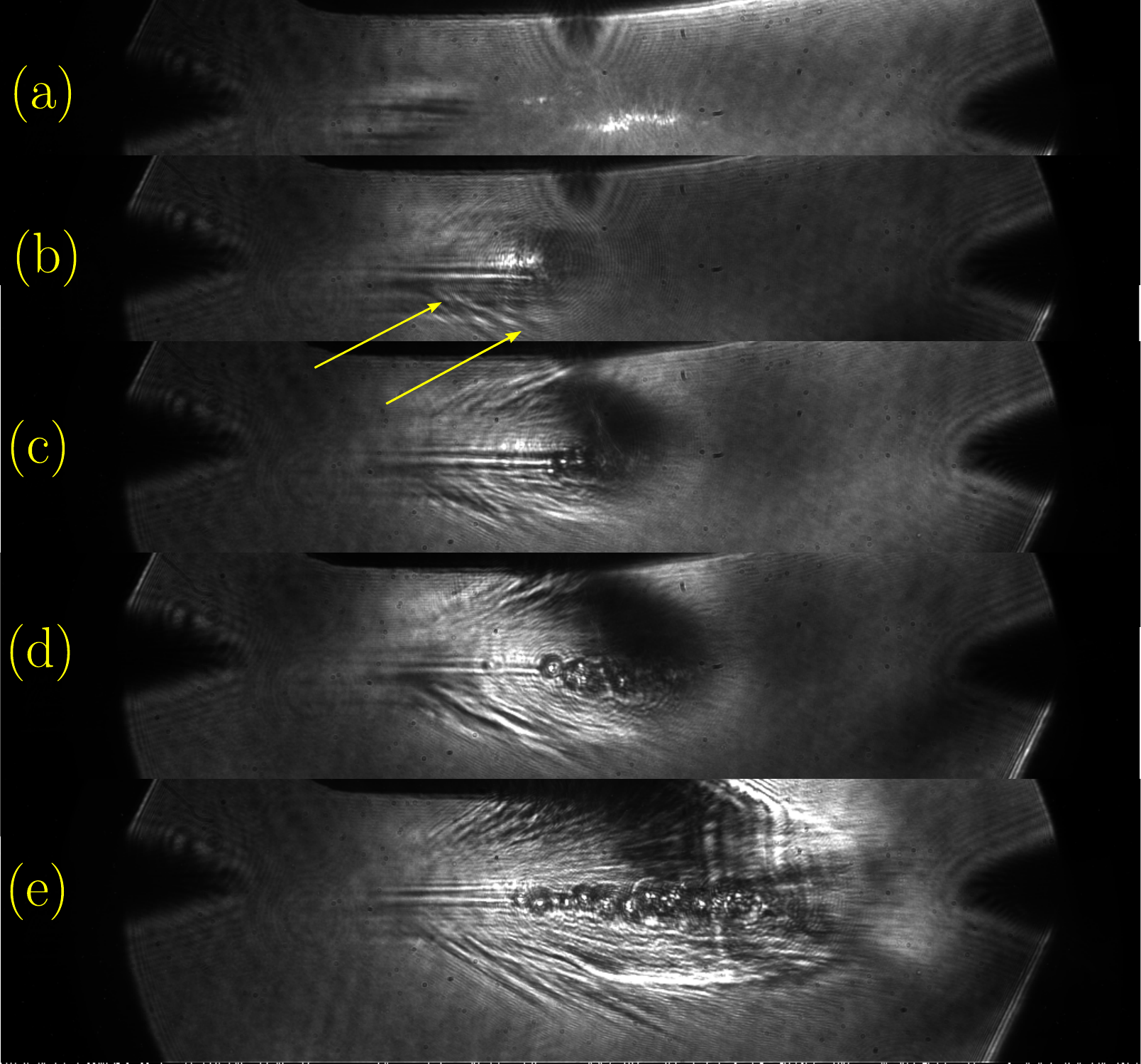}
\caption{Time-resolved pulse penetration in the plasma gradient. (a) reference instant
$t_0$, (b) $t_0+540$~fs, (c) $t_0+940$~fs, (d) $t_0+1.54$~ps, (e) $t_0+2.84$~ps. Arrows indicate the
growth of prominent laser filaments}.
\label{filamentation}
\end{figure}
\paragraph*{}
Besides being copiously generated by the interaction, the nonlinear structures
distribute spatially into two remarkable patterns, identifiable by following the bright central
blobs, which arose randomly in our experiment. Figure~\ref{hosing} illustrates both antisymetric
(zig-zag about the laser axis) (panel (a)) and axial (panel (b)) distributions of the
bubbles. In Ref.~\onlinecite{chen07}, the authors carried out 2D particle-in-cell simulation with $a_0=3$, $\tau=30$~fs and about $1$ mm of plasma at $n_e=10^{19}$ cm$^{-3}$. They obtained for
initially p- and s-polarized light (i.e. an electric field collinear and normal to the observation
plane), respectively antisymetric and axial density map patterns resembling our experimental results
(see figures~3 and 4 in \cite{chen07}). In particular, the antisymetric modulation is ascribed to the hosing
instability (see also \cite{naum01}). Although in our experiment the incident laser
polarization is normal to the observation plane, corresponding to the s-polarization of simulations
\cite{chen07,naum01}, hosing instabilities have also been observed experimentally in similar
conditions \cite{kalu10a}. Therefore, following \cite{chen07,naum01,kalu10a}, we propose to
ascribe our observations
to the laser excitation of the hosing instability inducing the developpement of von Karm\`an row of electron
vortices (a) and to shadowgraphic traces left by post-solitons (b).
\paragraph*{}
At that point, it is clear that the interaction in this operating regime ($n_e\sim10^{20}$ cm$^{-3}$,
long gradient $\sim250~\mu$m)
produces numerous instabilities (filamentation, hosing, vortex/soliton), known to seriously deplete the
pulse energy \cite{esir02,naum01,bula99}, and prevent efficient plasma channel charging for subsequent
Coulomb explosion. Moreover, despite their expansions by ion core explosion
\cite{naum01,borg02,sanc11a}, the bubbles do not provide a detectable acceleration, possibly because
of a number of accelerated ions below the detector noise level. Therefore, we were unable to verify
the post-soliton acceleration scenario predicted in \cite{esir02,sanc11a}.   
\begin{figure}[t]
\includegraphics[scale=0.6]{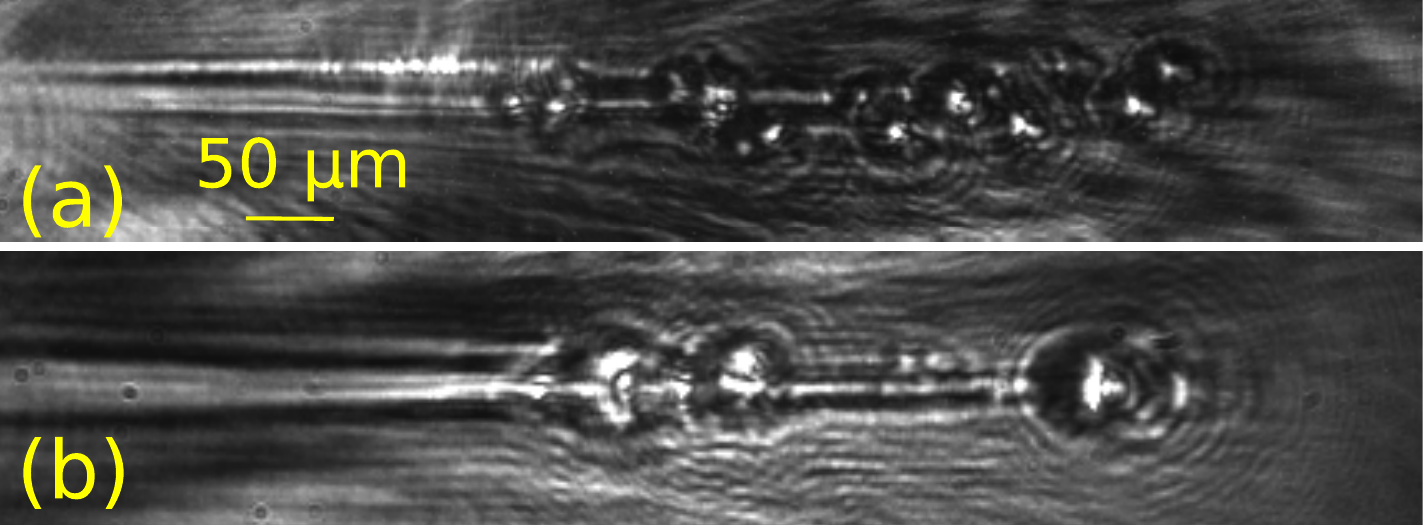}
\caption{Spatial distribution of the bubbles for two shots in the same conditions at $t_0+1.54$~ps
showing (a) antisymetric (zig-zag) and (b) symetric axial patterns about the main plasma channel}
\label{hosing}
\end{figure}
\paragraph*{}
To confirm the electromagnetic nature of the coherent structures, we carried out
some polametric measurements. To simplify these measurements, we have identified a regime where few
structures are generated at each laser shot, near the center of the jet (statistically close to one). For that, we used a stretched laser
pulse $\tau=250$~fs carrying the same energy ($a_0=0.8$) and a small
sonic nozzle of $200~\mu$m diameter delivering a jet of peak density of few 10$^{19}$
cm$^{-3}$ which we let expand in vacuum to increase the gradient before the interaction. This setup enables $(i)$ a well-localized interaction, $(ii)$ smoother interferograms and polarigrams without
filaments and bubble overlapping $(iii)$ efficient energy transfer to the structures \cite{naum01}.
\begin{figure}[h]
\includegraphics[scale=0.8]{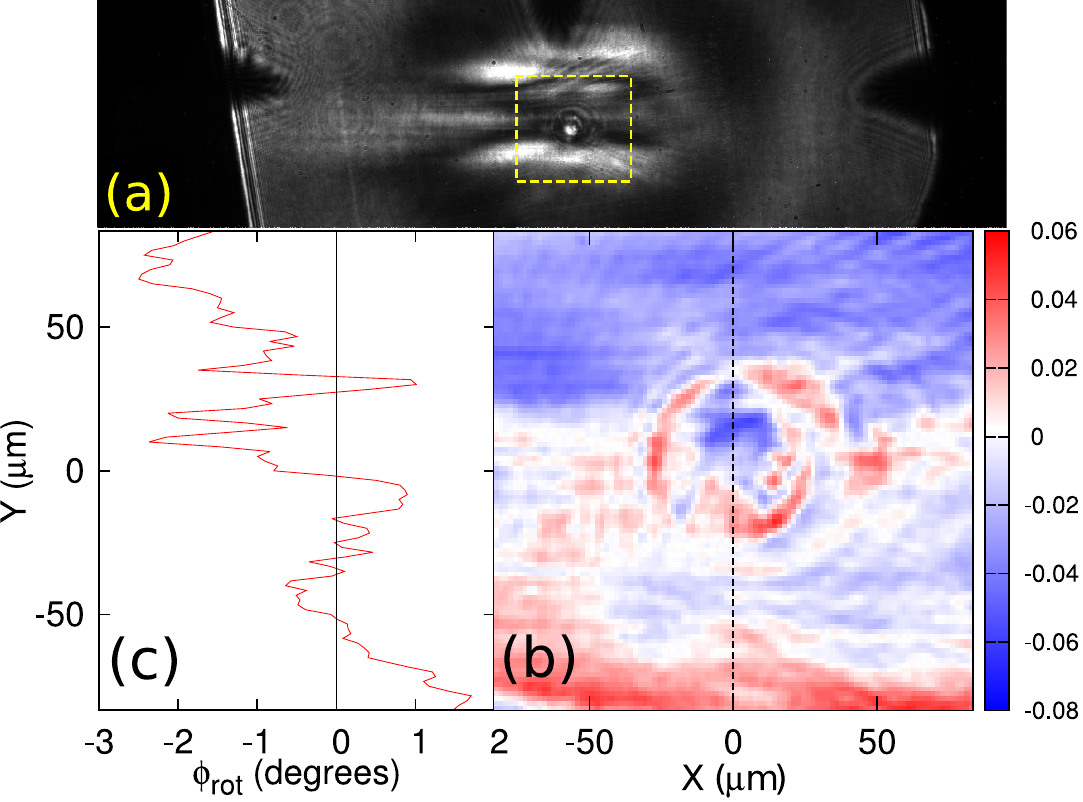}
\caption{(a) Shadowgram with one isolated structure in the laser wake at $t_0+940$~fs
($\tau=350$~fs, 10$^{19}$ cm$^{-3}$). (b) $\phi_{rot}$ map (in radians) of the yellow square dotted zone in (a). Detuning angle of the polarisers:
$\pm 10^{\circ}$. (c) Line-out along the black dotted line in (b)}
\label{bmag}
\end{figure}
\paragraph*{}
Figure~\ref{bmag}(a) shows a shadowgram with one single spherical structure of $57~\mu$m diameter in a clean area. In that
simplified case, we were able to extract a $\phi_{rot}$ map around the structure (panel
(b)). From panel (b) and (c), we can deduce that the structure in this condition has an almost
axisymmetric polarimetric imprint about the laser polarization direction. The density in the bubble shell is found to be $\sim7\times10^{19}$
cm$^{-3}$, five times that of the surrounding plasma. Though the polarimetric information is the
convolution of $B_z$ by the density along
the integrated probe beam path in the plasma, we will assume for the remaining that main contribution comes from the structure and
not from the surrounding plasma, as the major part of the bubble seems to be lying in a weakly
magnetized area (white region). Doing so, we thus calculate the maximum magnetic field
extractable from our data. For $n_e=7\times10^{19}$ cm$^{-3}$, an average path $l_z$ in the shell of
$25~\mu$m (from panel (c)), $n_c=4\times1.67\times10^{21}$ cm$^{-3}$ and a Faraday rotation of 0.04 radians, we find
$B_{z}\leq\frac{\phi_{rot}}{\frac{n_e l e}{2m_ecn_c}}=570$ T. Though lower, this value is of the
same order of magnitude as in the numerical computations in similar conditions \cite{bula99,esir02} and
is consistent with the fact that the fields dissipate as the structure expands. The extraction
of the space-resolved magnetic field map and a refined magnetic description of the nonlinear
structures will be the subject of another publication.
\paragraph*{}
In conclusion, our reproducible observations of bubbles identified as nonlinear coherent structures
(vortex/post-soliton) made it possible to demonstrate their detrimental effect on the efficiency of
laser-ion accelaration. These structures appear abundantly in the laser wake when the plasma has a
long gradient profile in which the laser self-focuses, filaments and develop hosing. When using a
small nozzle with lower density, simple conditions with one structure per laser shot can be
statistically obtained. We think this all-optical setup represents a powerful method that opens up new opportunity of in-depth investigation
of fundamental nonlinear plasma instabilities and also particle acceleration for users of
well-spread TeraWatt-class lasers.
\paragraph*{}
The authors thank Prof. S. V. Bulanov for fruitful and encouraging discussions. We are also pleased
to acknowledge the support of the European Research Council for funding the PARIS
ERC project (contract 226424), and the ANR-08-NT08-1-38025-1. Finally, the authors acknowledge the support of ANR under the GOSPEL project, grant reference ANR-08-BLAN-0072-03.
\bibliography{fran_0611}
\end{document}